\documentclass{article}

\usepackage{microtype}
\usepackage{graphicx}
\usepackage{subfigure}
\usepackage{booktabs} 

\usepackage{hyperref}

\usepackage[accepted]{icml2020}

\usepackage{times}
\usepackage{epsfig}
\usepackage{amsmath}
\usepackage{amssymb}
\usepackage{algorithm}
\usepackage{algorithmic}
\usepackage{stfloats}
\usepackage[T1]{fontenc}
\usepackage[utf8]{inputenc}
\usepackage{multirow}

\icmltitlerunning{Residual-Recursion Autoencoder for Shape Illustration Images}

\begin{document}

\twocolumn[
\icmltitle{Residual-Recursion Autoencoder for Shape Illustration Images}



\icmlsetsymbol{equal}{*}

\begin{icmlauthorlist}
\icmlauthor{Qianwei Zhou}{zjut}
\icmlauthor{Peng Tao}{zjut}
\icmlauthor{Xiaoxin Li}{zjut}
\icmlauthor{Shengyong Chen}{tjut}
\icmlauthor{Fan Zhang}{zjut}
\icmlauthor{Haigen Hu}{zjut}
\end{icmlauthorlist}

\icmlaffiliation{zjut}{College of Computer Science and Technology, Zhejiang University of Technology, Hangzhou, China. zhouqianweischolar@gmail.com, jstaopeng@qq.com, mordekai@zjut.edu.cn, zf@zjut.edu.cn, hghu@zjut.edu.cn.}
\icmlaffiliation{tjut}{School of Computer Science and Engineering, Tianjin University of Technology, Tianjin, China. sy@ieee.org.}

\icmlcorrespondingauthor{Haigen Hu}{hghu@zjut.edu.cn}

\icmlkeywords{Autoencoder, Deep Learning, Shape Illustrator, Gray Image, Binary Image}

\vskip 0.3in
]



\printAffiliationsAndNotice{}  

\begin{abstract}
Shape illustration images (SIIs) are common and important in describing the cross-sections of industrial products. Same as MNIST, the handwritten digit images, SIIs are gray or binary and containing shapes that are surrounded by large areas of blanks. In this work, Residual-Recursion Autoencoder (RRAE) has been proposed to extract low-dimensional features from SIIs while maintaining reconstruction accuracy as high as possible. RRAE will try to reconstruct the original image several times and recursively fill the latest residual image to the reserved channel of the encoder's input before the next trial of reconstruction. As a kind of neural network training framework, RRAE can wrap over other autoencoders and increase their performance. From experiment results, the reconstruction loss is decreased by 86.47\% for convolutional autoencoder with high-resolution SIIs, 10.77\% for variational autoencoder and 8.06\% for conditional variational autoencoder with MNIST.
\end{abstract}

\section{Introduction}
\begin{figure}[htb]
\begin{center}
\includegraphics[width=1.0\linewidth]{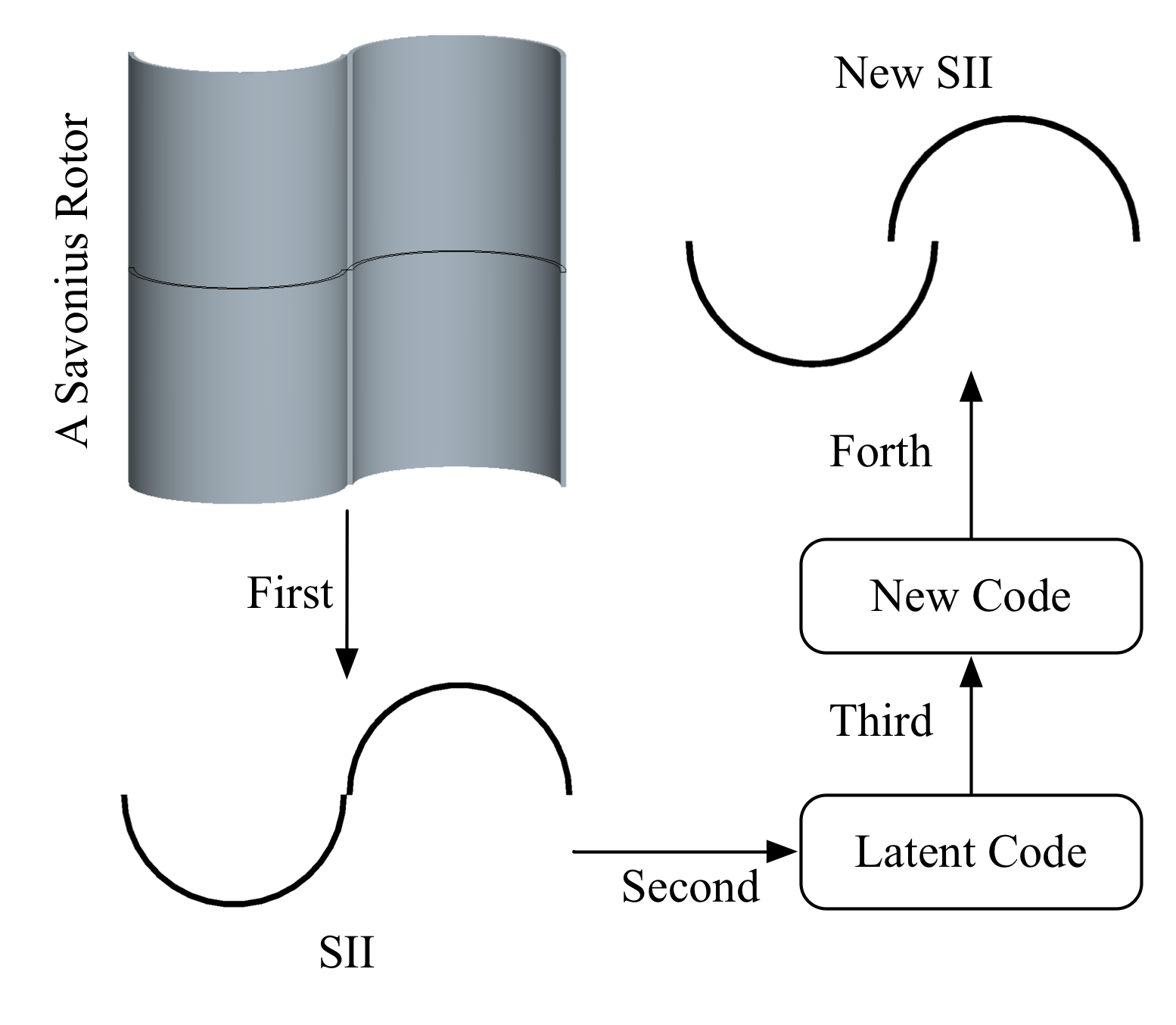}
\end{center}
   \caption{An example of the SII-based CAD method~\cite{zhou2018innovative} . First, use the cross section as the SII to describe the rotor. Second, compress the SII by 2D discrete cosine transformation. Third, use a Genetic Algorithm to generate a new code from the old one. Forth, use inverse 2D discrete cosine transformation to reconstruct the new SII.}
\label{fig:SIIexp}
\end{figure}

Recently, \cite{zhou2018innovative} proposed a method for computer-aided design (CAD) which can find efficient and innovative CAD models automatically. The method will first describe the target product by a shape illustration image (SII), Figure~\ref{fig:SIIexp} for example; second compress the SII to get a low dimensional latent code $z$; third modify $z$ by a random searching algorithm to get a new latent code $z_{new}$; fourth reconstruct a new SII from $z_{new}$ and test its performance by a computer-aided engineering software. A highly efficient SII will be found automatically by replacing $z$ with $z_{new}$ and doing the third and the fourth steps repeatedly. The image encoding and decoding techniques are the bottlenecks of the CAD method. A high compression rate can get low dimensional $z$ that will lead to shorter optimization periods. But current encoding techniques are all sacrificing details to get a high compression rate which is unacceptable to SIIs whose details are highly correlated to their performance. 

Because SIIs are usually generated by following predefined principals, they are similar to each other even though they are containing sharp edges and/or lots of small shapes. It is possible to learn the principals by deep learning-based autoencoders and express them with low dimensional features. Most autoencoders~\cite{chen2016variational,nalisnick2016stick,xu2019stacked,qi2014robust,dong2018review,bojanowski2017optimizing,sonderby2016ladder,wang2012folded,creswell2018denoising,kiasari2018coupled,wang2016auto} are using the traditional structure where images are encoded into low dimensional features and then decoded into the reconstructed images. When dealing with SIIs, the traditional autoencoders are shot-handed because of the lack of mechanics to emphasize hard patterns. The hard patterns are part of details that distinguish an SII from the others and are difficult to be captured by autoencoders. \cite{zhao2018unsupervised} proposed to learn features with image pyramids generated by smoothing and down-sampling operations. Although image pyramids can highlight details, the details are found by non-trainable operations that are not necessarily capable of identifying the hard patterns.  

In this work, a framework, namely Residual-Recursion Autoencoder (RRAE), has been proposed to encode SIIs into low dimensional latent code recursively. RRAE will try to reconstruct the original image $T$ times. The input of the autoencoder has $T$ channels whose first channel is the original image. The residual between the reconstructed image and the original image will be filled to its reserved channel in the input. The updated input will be used to encode and reconstruct the original image again. At $T$th autoencoder forward propagation, the output of the encoder will be kept as the latent code and the decoder output will be the final reconstructed image. By the residual-recursion mechanic, the hard patterns are detected by a trainable operator, the autoencoder itself. The hard patterns will be highlighted in each channel of the input except the first channel. RRAE can wrap over different autoencoders and increase their performance. From the experiment results, the reconstruction loss is decreased by 86.47\% for convolutional autoencoder with high-resolution SIIs, 10.77\% for variational autoencoder and 8.06\% for conditional variational autoencoder with MNIST. Because high resolution means more hard patterns, autoencoders have been improved by big margins on high-resolution SIIs.

\section{Methodology}
\begin{figure}[htb]
\begin{center}
\includegraphics[width=1.0\linewidth]{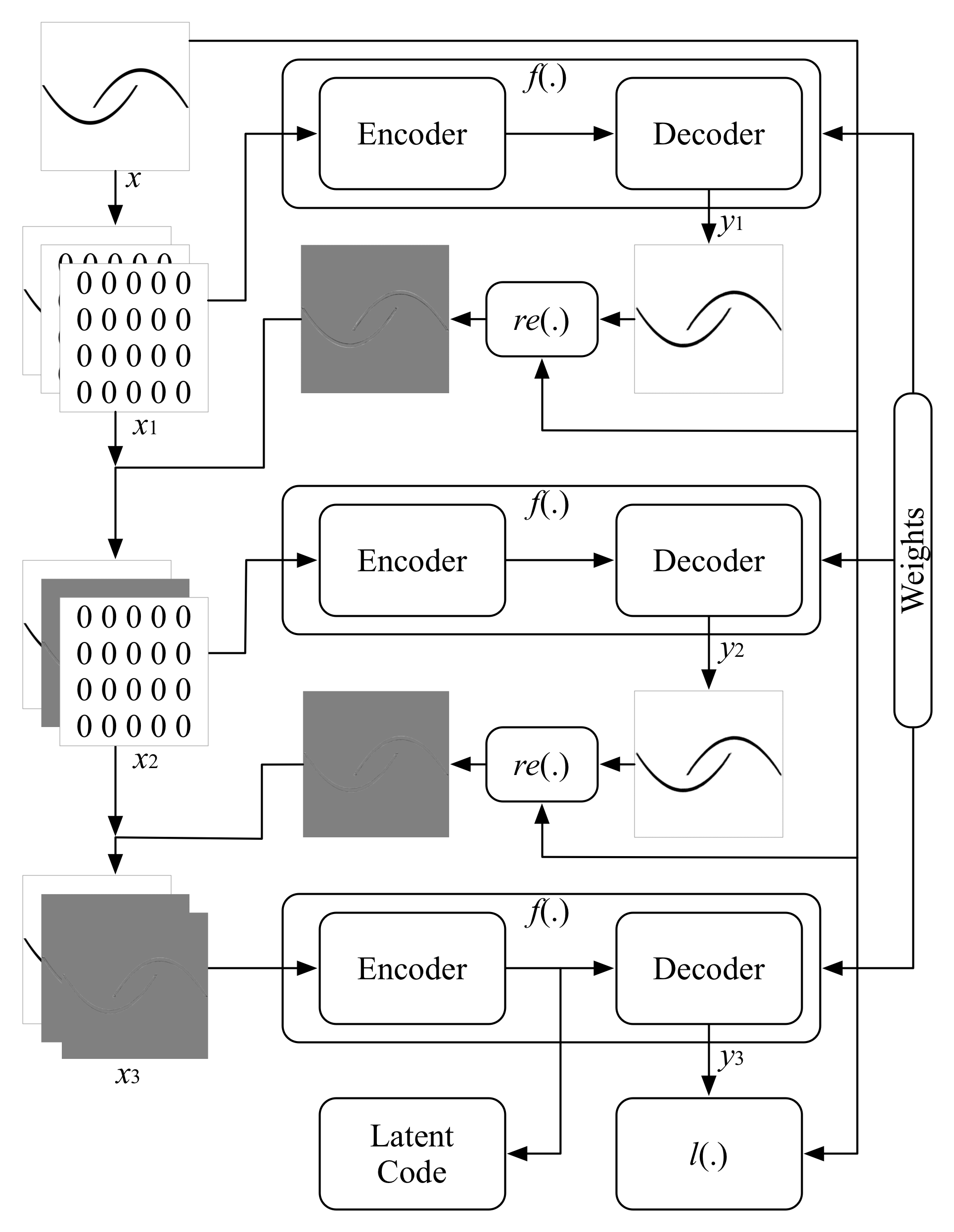}
\end{center}
   \caption{An example of the RRAE algorithm. $X$ is an SII dataset consisted of cross-sectional images of Savonius rotors. Three autoencoders $f(.)$s share the same set of weights. The residual function is $re(x,y_t)=(x-y_t)/2+0.5$ where pixel value is range from 0 to 1. The loss function $l(.)$ is the L1 loss of $x$ and $y_3$.}
\label{fig:wf}
\end{figure}

Algorithm~\ref{alg:RRAE} shows the Residual-Recursion Autoencoder (RRAE). The autoencoder network $f(.)$ can be any structure that takes a tensor as input and outputs a reconstructed one. The loss function $l(.)$ consists of loss functions that are required by $f(.)$. For example, $l(.)$ can be $l(x,y,z)=\|x-y\|+\|z\|$ where $x$ is the input image, $y$ is the reconstructed image and $z$ is the latent code. Minimizing $l(x,y,z)$ will minimize reconstruction error and impose the sparsity of the latent code. There is no limitation on the optimization function $o(.)$ as long as it can work with $f(.)$ and optimize its weights $\theta$. The residual function $re(.)$ is used to get the residual. For example, $re(x,y)=(x-y)/2$. Supposing $T=3$, Figure~\ref{fig:wf} shows an example.
 
\begin{algorithm}  
\scriptsize
  \renewcommand{\algorithmicrequire}{\textbf{Input:}}
  \renewcommand{\algorithmicensure}{\textbf{Output:}}
  \caption{Residual-Recursion Autoencoder}
  \label{alg:RRAE}
  \begin{algorithmic}
    \REQUIRE Dataset $X$, autoencoder network $f(.)$ and its weights $\theta$, total times of reconstruction trial $T$, iteration times $N$, loss function $l(.)$ and optimization function $o(.)$, residual function $re(.)$.
    \ENSURE The optimal set of weights $\theta^*$ which minimizes $l(.)$. 
    \FOR{$i=1$ {\bfseries to} $N$}
    \STATE  from $X$ get an image randomly as $x$.
    \FOR{$t=1$ {\bfseries to} $T-1$}
    \STATE  $x_t=[r_0,r_1,\dots,r_{t-1},0,\dots,0]$, where $r_0=x$, $x_t$ is a $T$ channels tensor.
    \STATE  $y_t,z_t=f(x_t,\theta)$, where $y_t$ is the reconstructed image and $z_t$ is the latent code.
    \STATE  $r_t = re(x,y_t)$.
    \ENDFOR 
    \STATE $x_T=[y_0,y_1,\dots,y_{T-1}]$
    \STATE $y_T,z_T=f(x_t,\theta)$
    \STATE $\theta=o(l(x,y_T,z_T),\theta)$
    \ENDFOR
    \STATE \textbf{return} $\theta^*$.
  \end{algorithmic}  
\end{algorithm}

\section{Experiments}
All experiments have run on 1080ti GPU and Pytorch~\cite{paszke2019pytorch} framework. As default, the optimization function is Adam~\cite{kingma2014adam} with default configurations, learning rate is 1e-4, weight decay is 1e-5, the residual function is $re(x,y)=(x-y)/2$, total epoch number is 300.
\subsection{MNIST}
MNIST~\cite{lecun1998gradient} is a handwritten digital dataset in which 60000 images for training and 10000 for testing. Images of MNIST are similar to SIIs except the resolution is much lower than SIIs'. In this experiment, code from Github~\footnote{https://github.com/timbmg/VAE-CVAE-MNIST} has been modified to run RRAE on MNIST with Variational Autoencoder (VAE)~\cite{kingma2013auto} and Conditional Variational Autoencoder (CVAE)~\cite{sohn2015learning} as the autoencoder $f(.)$ respectively whose networks are consisted of Linear and ReLU layers. The encoding and decoding part is wrapped by RRAE where the autoencoder will try reconstruct the image $T$ times and return the reconstructed image $y_T$, the latent code mean $z_{\mu}$, the natural logarithm of latent code variance $z_{\ln\sigma^2}$ and the latent code $z_T$ of the last trial. Before every trial, the latest residual will be filled to $x_t$ where $re(x,y_{t-1})=(x-y_{t-1})/2$ which also is the default residual function of all experiments in this work~\footnote{Please refer to the uploaded code in VAE-CVAE folder for implementation details.}. According to the code, the loss function is $l(x,y_T,z_{\mu},z_{\ln\sigma^2})=BCE(x,y_T)+KLD(z_{\mu},z_{\ln\sigma^2})$, where $BCE(.)$ and $KLD(.)$ are Binary Cross Entropy (equation~\eqref{eq:BCE})  and KL divergence (equation~\eqref{eq:KLD}). $\sum$ and other operations are all pixel- or element-wise. The results are listed in Table~\ref{tab:VAE} where total epoch number is 300, learning rate is 0.001 without weight decay, the optimization function $o(.)$ is Adam~\cite{kingma2014adam}, the first column is the dimension of the latent code $z$, the second column is the total trial times $T$, the MSE column is the best testing mean square error between original image and reconstructed image during training, the DR column is the decrease rate of MSE respecting to the baseline (the $T=1$ result). From Table~\ref{tab:VAE}, it is obvious that the RRAE helps a lot in decreasing reconstruction error without increasing the dimension of the latent code. Usually, bigger $T$ leads to better performance. But too many trials will consume too much computation with little improvements. So, in the following experiments, the upper limit of $T$ is 3.

\begin{align}
BCE(x,y) & = -\sum(x\ln(y)+(1-x)\ln(1-y))
\label{eq:BCE}
\end{align}
\begin{align}
KLD(m,n) & = -0.5\sum(1+n-m^2-\exp(n))
\label{eq:KLD}
\end{align}

\begin{table}[htb]
\centering
\caption{VAE and CVAE experiments.}
\resizebox{0.5\textwidth}{!}{
\begin{tabular}{cccccc}
\hline
\multirow{2}{*}{$z$} & \multirow{2}{*}{$T$} & \multicolumn{2}{c}{VAE} & \multicolumn{2}{c}{CVAE} \\ 
                   &                    & MSE          & DR/\%    & MSE        & DR/\%       \\ \hline
2                  & 1                  & 0.039362     & 0        & 0.033084   & 0           \\
2                  & 2                  & 0.037122     & 5.69     & 0.031937   & 3.47        \\
2                  & 3                  & 0.036552     & 7.14     & 0.031546   & 4.65        \\
2                  & 4                  & 0.037878     & 3.77     & 0.031456   & 4.92        \\ \hline
5                  & 1                  & 0.025071     & 0        & 0.021242   & 0           \\
5                  & 2                  & 0.024199     & 3.48     & 0.020755   & 2.29        \\
5                  & 3                  & 0.023470     & 10.77    & 0.020561   & 3.21        \\
5                  & 4                  & 0.023040     & 8.1      & 0.020649   & 2.79        \\ \hline
10                 & 1                  & 0.015637     & 0        & 0.013968   & 0           \\
10                 & 2                  & 0.014536     & 7.04     & 0.013009   & 6.87        \\
10                 & 3                  & 0.014621     & 6.50     & 0.012842   & 8.06        \\
10                 & 4                  & 0.014175     & 9.35     & 0.012852   & 7.99        \\	\hline
\end{tabular}}
\label{tab:VAE} 
\end{table}

Convolutional autoencoders have been tested on MNIST and its high-resolution version~\footnote{Please refer to code in folder CNN for implementation details.}. The autoencoders are piled up by layers of  2D convolution, Group Normalization~\cite{wu2018group} and ReLU without skipping links. The high-resolution MNIST is a 512x512 binary image dataset that is generated by bilinear interpolation in which 60000 images for trianing and 10000 for testing. The images are binarized with threshold 127.5. All images are mean and std normalized. The loss function $l(.)$ of the original MNIST is L1. For high resolution, the loss function is $l(.)=NMS(.)=(1-MS(.))*100$~\cite{mentzer2018conditional} where $MS(.)$ is MS-SSIM~\cite{wang2003multiscale}. Table~\ref{tab:CAE} shows the best testing results from which we can conclude that the RRAE is much more efficient for high-resolution images than low-resolution images.  

\begin{table}[htb]
\centering
\caption{Convolutional autoencoders with MNIST and its high-resolution version.}
\resizebox{0.5\textwidth}{!}{
\begin{tabular}{cccc|cccc}
\hline
\multicolumn{4}{c}{MNIST28} & \multicolumn{4}{|c}{MNIST512} \\
$z$  & $T$  & L1  & DR/\%  & $z$    & $T$  & $NMS(.)$  & DR/\%  \\ \hline
1  & 1  & 0.07645  & 0      & 50   & 1  & 2.851    & 0      \\
1  & 2  & 0.07895  & -3.27  & 50   & 2  & 1.800    & 36.86  \\ \cline{1-4}
2  & 1  & 0.06544  & 0      & 50   & 3  & 1.534    & 46.19  \\ \cline{5-8}
2  & 2  & 0.06358  & 2.84   & 100  & 1  & 1.399    & 0      \\
2  & 3  & 0.06425  & 1.82   & 100  & 2  & 0.4281   & 69.40 \\ \hline
\end{tabular}}
\label{tab:CAE} 
\end{table}

\subsection{SIIs of Savonius Rotors}
\begin{figure*}[htb]
\begin{center}
\includegraphics[width=1.0\linewidth]{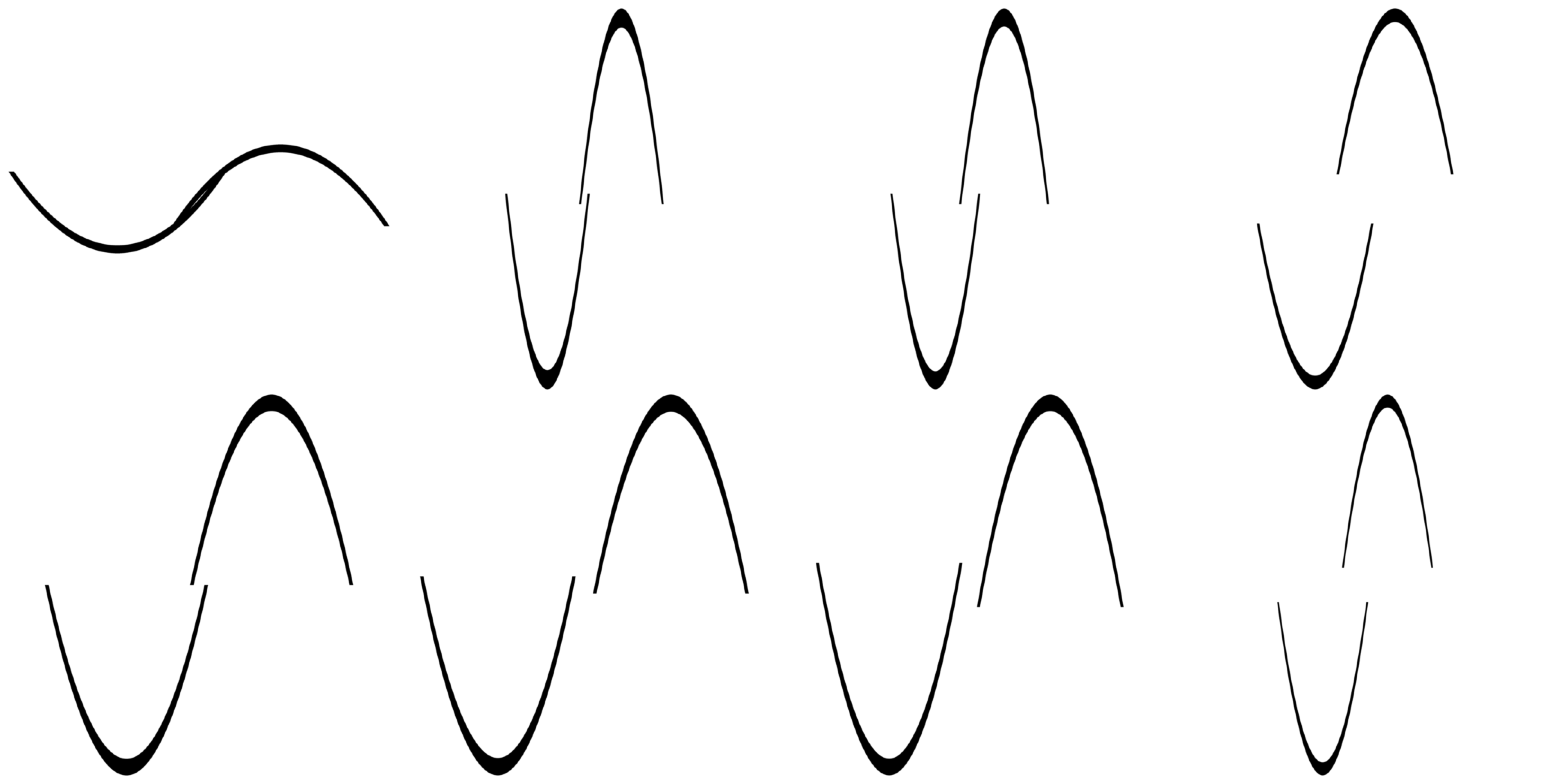}
\end{center}
   \caption{Random samples of the SII dataset.}
\label{fig:SIIdataset}
\end{figure*}

A SII dataset has been constructed according to~\cite{zhou2018innovative} which is consisted of cross-sectional images of Savonius Rotors. Since the shape of a Savonius Rotor is controlled by a parabolic curve that is specified by four enumerable variables, 26973 SIIs have been generated by enumerating the height $h_1=100,125,\dots,1000$, the length $l=400,425,\dots,600$, the down left point $(x_1,y_1)$ where $x_1=-100,-75,\dots,100$ and $y_1=-100,-75,\dots,100$~\footnote{Please refer to code /CNN/parabolarBlade.py for detailed implementation.}. Figure~\ref{fig:SIIdataset} shows some random samples of the SII dataset. The resolution is 512x512. 1 out of 7 images are selected randomly for testing. The others are kept for training. 

\begin{figure*}[htb]
\begin{center}
\includegraphics[width=1.0\linewidth]{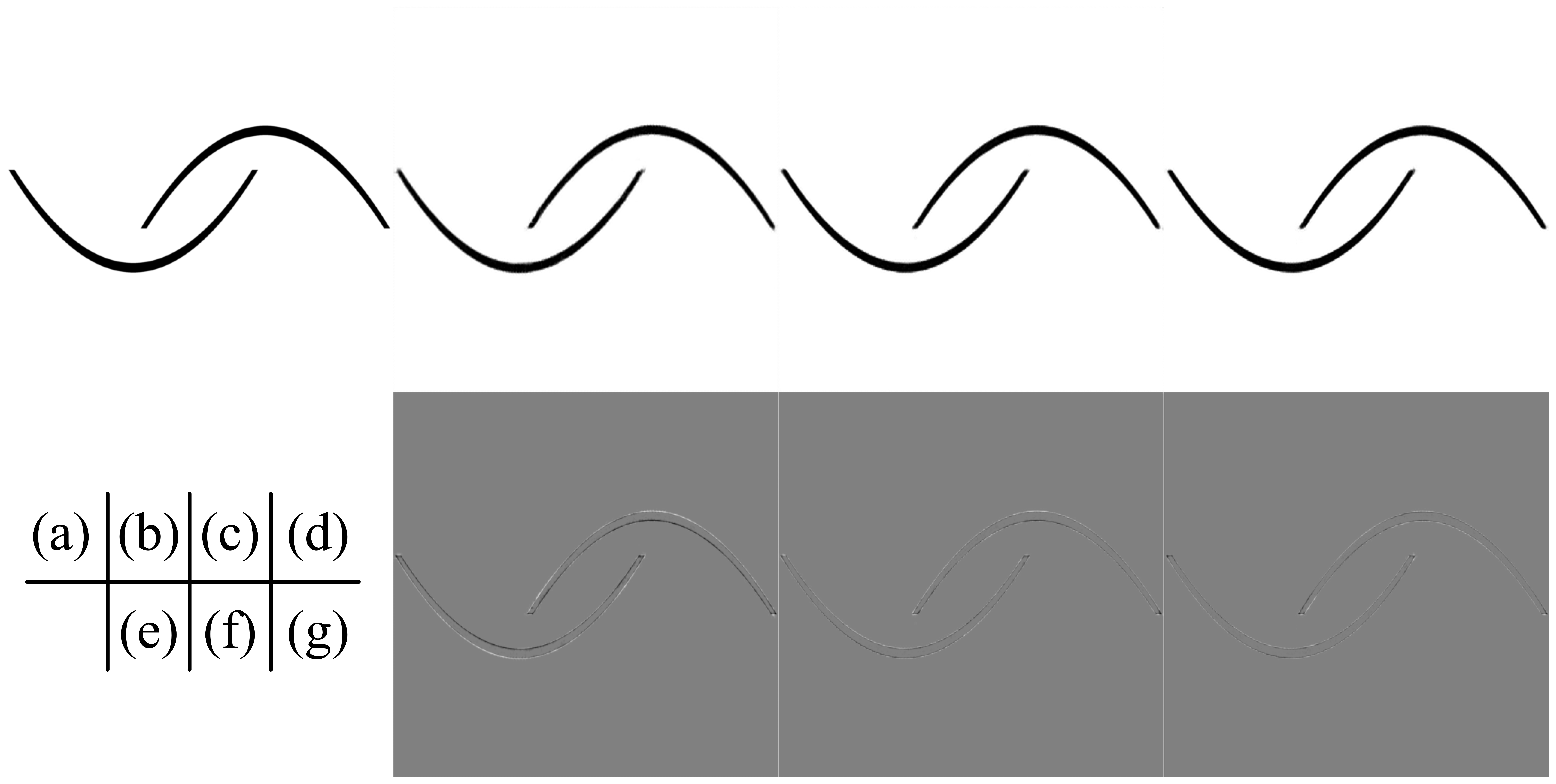}
\end{center}
   \caption{Results of each trial. (a) is the original image. (b) is the reconstructed image of the first trial. (c) and (d) are the reconstructed images of the second and third trials. (e) is the residual image of the first trial. (f) and (g) are the residual images of the second and third trials.}
\label{fig:ret}
\end{figure*}

A convolutional autoencoder that wrapped by RRAE is used to encode the cross-sectional images~\footnote{Please refer to code in folder CNN for implementation details.}. The autoencoder is piled up by layers of  2D convolution, batch normalization~\cite{ioffe2015batch} and ReLU without skipping links. The input cross-sectional image is mean and std normalized. The output of the last transposed convolution layer is denormalized to get the reconstructed image. The reconstructed image and the original cross-sectional image are used to calculate the residual image and the $NMS(.)$ loss. Table~\ref{tab:csire} shows the results of the cross-sectional autoencoding experiment in which the best test results have been listed. From the results, RRAE has improved the reconstruction accuracy significantly. When the latent code dimension is low, 2 for example in this experiment, too many trials (e.g. $T=3$) will harm the performance of RRAE. The performance decreasing can also be observed in Table~\ref{tab:CAE}. Figure~\ref{fig:ret} shows the results of each trial that are from one test run of the $z=5$ \& $T=3$ experiment. The reconstructed images are clamped to the range from 0 to 1. To illustrate the residual images, they are added by the bias 0.5 to keep the pixel value in the range from 0 to 1. From Figure~\ref{fig:ret}, it is obvious that the residual gets smaller after each trial. The final reconstructed image ($t=T$) is very close to the original image which has been encoded into just 5 float variables. 

\begin{table}[htb]
\centering
\caption{The cross-sectional image autoencoding experiment.}
\resizebox{0.5\textwidth}{!}{
\begin{tabular}{cccc|cccc}
\hline
$z$  & $T$  & $NMS(.)$  	& DR/\%  & $z$    & $T$  & $NMS(.)$  & DR/\%  \\ \hline
2  & 1  & 5.069   & 0      	& 4	   	& 1  & 0.2821   	& 0      \\
2  & 2  & 3.665   & 27.70  & 4	   	& 2  & 0.05777    	& 79.52  \\ 
2  & 3  & 4.910   & 3.14    & 4	   	& 3  & 0.04142    	& 85.32  \\ \hline
3  & 1  & 0.7347 & 0   		& 5	 	& 1  & 0.2081    	& 0      \\
3  & 2  & 0.2983 & 59.40	& 5		& 2  & 0.04240   	& 79.63 \\ 
3  & 3  & 0.2591 & 65.73	& 5		& 3  & 0.02815  	& 86.47 \\ \hline
\end{tabular}}
\label{tab:csire} 
\end{table}

\begin{figure*}[htb]
\begin{center}
\subfigure[DCT]{  
\begin{minipage}[b]{0.31\linewidth}
\centering
\includegraphics[width=1\linewidth]{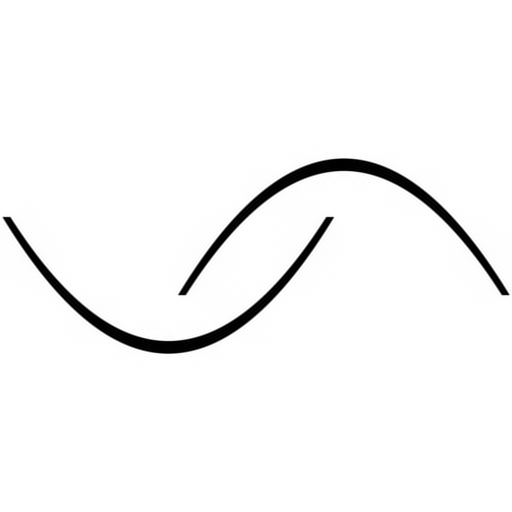} \\
\vspace{1pt}
\includegraphics[width=1\linewidth]{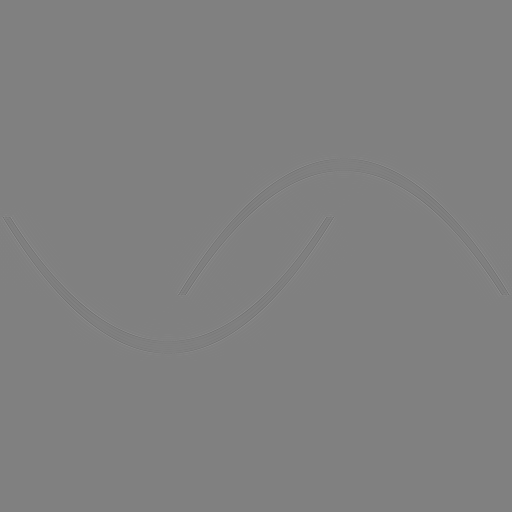}
\end{minipage}
}
\hspace{0.05cm}    
\subfigure[Jpeg]{  
\begin{minipage}[b]{0.31\linewidth}
\centering
\includegraphics[width=1\linewidth]{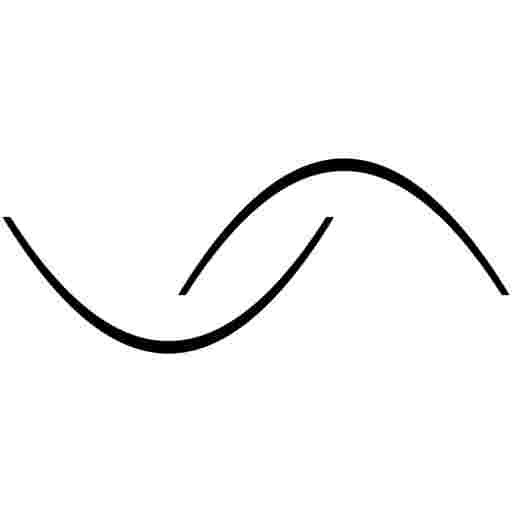} \\
\vspace{1pt}
\includegraphics[width=1\linewidth]{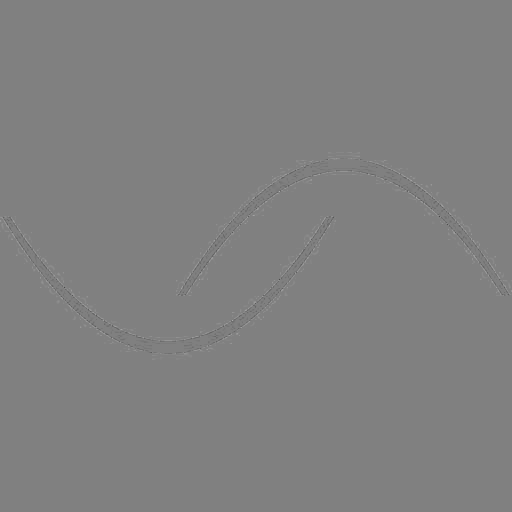}
\end{minipage}
}
\hspace{0.05cm}
\subfigure[Jpeg2000]{  
\begin{minipage}[b]{0.31\linewidth}
\centering
\includegraphics[width=1\linewidth]{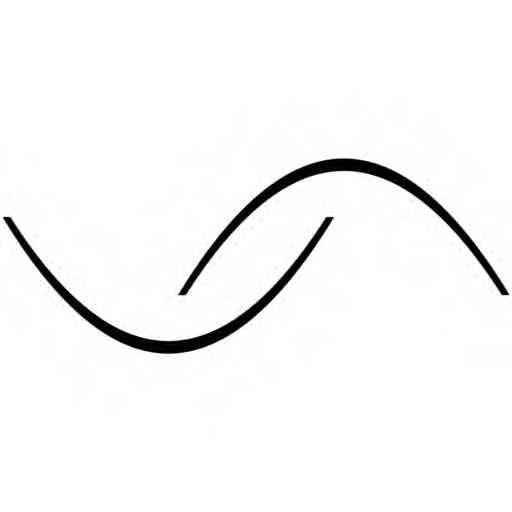} \\
\vspace{1pt}
\includegraphics[width=1\linewidth]{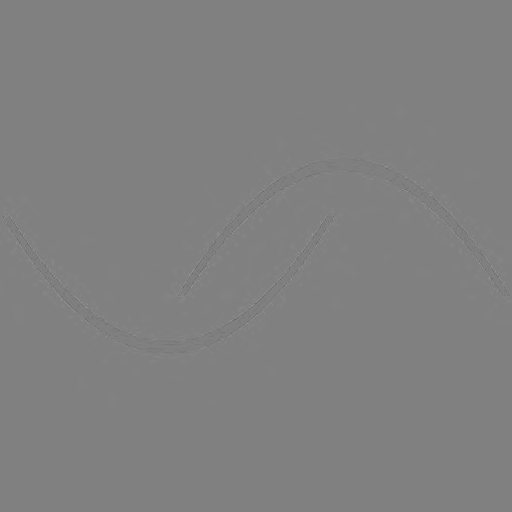}
\end{minipage}
}
\end{center}
   \caption{Image compression results.}
\label{fig:dct}
\end{figure*}

The L1 loss of the final reconstructed image is 0.0020 in Figure~\ref{fig:ret}. Figure~\ref{fig:dct} shows the results of different image compression algorithm with the same original image. The DCT method is same as the one introduced in~\cite{zhou2018innovative} where a image will be encoded into latent codes by 2D discrete cosine transformation and Zigzag reordering~\footnote{Please refer to code in folder DCT for implementation details.}. Jpeg and Jpeg2000 are provided by MATLAB2014a. The DCT method needs 119157 double variables to reconstruct the image with L1 loss 0.0020. Jpeg has the loss 0.0024 with file size 5177 bytes. Jpeg2000 has the loss 0.0021 with file size 3333 bytes. Although L1 losses are close to each other, the code length of image compression methods are much longer than RRAE. From the residual images of Figure~\ref{fig:dct}, there are obvious noises in the reconstructed images. Comparing to Figure~\ref{fig:dct}, the final reconstructed image in Figure~\ref{fig:ret} has smoother and cleaner edges that are important in describing CAD shapes.

\section{Conclusion}
An autoencoder framework, Residual-Recursion Autoencoder (RRAE), has been proposed to boost the performance of any autoencoder that encodes the target image into a latent code and reconstructs the image from the latent code. RRAE can endow the autoencoders with the ability of learning, capturing and highlighting hard patterns of the target image. When wrapped by RRAE, autoencoders will try to reconstruct the target image several times. After each trial, the residual between the reconstructed image and the target image will be filled to the reserved channel of the input tensor. Recursively, the input tensor will be full of residual images in which hard patterns may be repeated several times. With the fully filled input tensor, autoencoder can reconstruct the target image accurately with low dimensional latent code. The significant improvements over the baseline autoencoders have verified the performance of RRAE.

The target image should contain lots of hard patterns, for example, shape illustration images that consist of binary or gray shapes with sharp edges and large areas of blanks. Otherwise, RRAE will not bring in any significant improvements. This conclusion is supported by the comparative experiments of MNIST and its high-resolution version. RRAE with the high-resolution MNIST that contains much more hard patterns yielded more significant improvement than the original MNIST.

Supposing the computational cost of an autoencoder is $O(n)$, the cost will be $T \cdot O(n)$ after the wrapping of RRAE. According to the experiment results, the upper limit of $T$ is 3. In our experiments, RRAE increased computation cost by 2 times and decreased the reconstruction error by 86.47\%.

%


\bibliography{egbib}
\bibliographystyle{icml2020}

\end{document}